\documentclass[fleqn,usenatbib]{mnras}

\usepackage{newtxtext,newtxmath}

\usepackage[T1]{fontenc}

\DeclareRobustCommand{\VAN}[3]{#2}
\let\VANthebibliography\thebibliography
\def\thebibliography{\DeclareRobustCommand{\VAN}[3]{##3}\VANthebibliography}

\usepackage{graphicx}	
\usepackage{multirow}
\usepackage{booktabs}
\usepackage{float}
\usepackage{subfigure}
\usepackage{amsmath}	
\usepackage{longtable}
\usepackage{CJKutf8, color}
\usepackage{float}
\usepackage[figuresright]{rotating}
\usepackage{threeparttable}   

\DeclareRobustCommand{\hlsout}{\bgroup\markoverwith{\textcolor{red}{\rule[.5ex]{2pt}{0.4pt}}}\ULon}

\DeclareRobustCommand{\lysout}{\bgroup\markoverwith{\textcolor{red}{\rule[.5ex]{2pt}{0.4pt}}}\ULon}

\title{The double-peaked type I X-ray bursts with different mass accretion rate and fuel composition}

\author[Song et al.]{
Liyu Song,$^{1}$
Helei Liu,$^{1}$\thanks{E-mail: heleiliu@xju.edu.cn}
Chunhua Zhu,$^{1}$
Guoqing Zhen,$^{1}$
Guoliang L\"{u}$^{2}$
and Renxin Xu$^{3,4}$
\\
$^{1}$School of Physical Science and Technology, Xinjiang University, Urumqi 830046, China  \\
$^{2}$Xinjiang Astronomical Observatory, Chinese Academy of Science, 150 Science 1-Street, Urumqi 830011, China\\
$^{3}$Department of Astronomy, Peking University, Beijing 100871, China\\
$^{4}$Kavli Institute for Astronomy and Astrophysics, Peking University, Beijing 100871, China
}

\date{Accepted XXX. Received YYY; in original form ZZZ}

\pubyear{2015}

\begin{document}
\label{firstpage}
\pagerange{\pageref{firstpage}--\pageref{lastpage}}
\maketitle

\begin{abstract}
Using the MESA code, we have carried out a detailed survey of the available parameter space for the double-peaked type I X-ray bursts. We find that the double-peaked structure appears at mass accretion rate $\dot{M}$ in the range of $\sim(4-8)\times10^{-10}\,M_{\odot}/{\rm yr}$ when metallicity $Z=0.01$, while in the range of $\sim(4-8)\times10^{-9}\,M_{\odot}/\rm{yr}$ when $Z=0.05$. Calculations of the metallicity impact suggest that the double peaks will disappear when $Z\lesssim0.005$ for $\dot{M}=5\times10^{-10}\,M_{\odot}/\rm{yr}$ and $Z\lesssim0.04$ for $\dot{M}=5\times10^{-9}\,M_{\odot}/\rm{yr}$. Besides, the impacts of base heating $Q_{\rm b}$, as well as nuclear reaction waiting points: $^{22}\rm{Mg}$, $^{26}\rm{Si}$, $^{30}\rm{S}$, $^{34}\rm{Ar}$, $^{56}{\rm Ni}$, $^{60}\rm Zn$, $^{64}\rm{Ge}$, $^{68}\rm{Se}$, $^{72}\rm{Kr}$ have been explored. The luminosity of the two peaks decreases as $Q_{\rm b}$ increases. $^{68}{\rm Se}(p,\gamma){^{69}{\rm Br}}$ is the most sensitive reaction, the double peaks disappear assuming that $^{56}{\rm Ni}(p,\gamma)^{57}{\rm Cu}$ and $^{64}{\rm Ge}(p,\gamma)^{65}{\rm As}$ reaction rates have been underestimated by a factor of 100 and the $^{22}{\rm Mg}(\alpha,p)^{25}{\rm Al}$ reaction rate has been overestimated by a factor of 100,  which indicates that $^{22}{\rm Mg}$, $^{56}{\rm Ni}$, $^{64}{\rm Ge}$, $^{68}{\rm Se}$ are possibly the most important nuclear waiting points impedance in the thermonuclear reaction flow to explain the double-peaked bursts. Comparisons to the double-peaked bursts from 4U 1636-53 and 4U 1730-22 suggest that the nuclear origins of double-peaked type I X-ray bursts are difficult to explain the observed larger peak times ($t_{\rm p,1}\gtrsim4\,{\rm s}$, $t_{\rm p,2}\gtrsim8\,{\rm s}$) and smaller peak ratio($r_{1,2}\lesssim0.5$). The composition of ashes from double-peaked bursts is very different from the single-peaked bursts especially for the heavier p-nuclei.

\end{abstract}




\section{Introduction}
In low-mass X-ray binaries (LMXBs), the unstable burning of accreted fuel on the surface of a neutron star (NS) triggers type I X-ray bursts ~\citep{lewin1993x,strohmayer2006,galloway2008}. The typical type I X-ray burst exhibits a rapid rise (1-5$\,\rm{s}$) and an exponential decay within 10-100$\,\rm{s}$, showing a single peak in the light curve profile. However, some observed double-peaked bursts have been found in several NS-LMXBs ~\citep{1989A&A...208..146P,bhattacharyya2006A,li2021}, and the origins of the double peaks feature can be categorized into two types: the instrumental origin caused by a photospheric radius expansion(PRE) and the astrophysical origin~\citep{bult2019}. For the former, the temperature of the photosphere temporarily shifts out of the instrument passband due to radius expansion~\citep{1983ApJ...267..315P}, which causes the dip in the observed X-ray rate, while the bolometric luminosity shows a single peak. For the latter, the double-peaked structure exists in the bolometric luminosity.

Non-PRE bursts with double-peaked structure have been detected in five bursters, e.g., 4U 1636-536~\citep{bhattacharyya2006A,li2021}, 4U 1608-52~\citep{1989A&A...208..146P, 2021ApJ...910...37G}, GX 17+2~\citep{kuulkers2002}, GRS 1741.9-2853~\citep{pike2021} and 4U 1730-22~\citep{2022ApJ...940...81B,chen2023}, although PRE bursts with double peak structures both in X-rays and bolometric luminosity have been observed from 4U 1608-52~\citep{jaisawal2019} and SAX J1808.4-3658~\citep{bult2019}.

Several theories have been proposed to explain the double-peaked bursts: e.g., the heat transport impedance from the other zone caused a dip in the light curve in a two-zone accreting model~\citep{1984A&A...134..123R}, the two steps generation/release of the thermonuclear energy~\citep{1985ApJ...299..487S, Fujimoto1988}, scattering from an accretion-disk corona~\citep{1987ApJ...315L..43M,melia1992}, thermonuclear flame spreading model (e.g., ignition at high latitude but stalls on the equator)~\citep{bhattacharyya2006A}. However, it was argued that the two-zone model is too coarse, the double peaks structure will disappear with increasing the zones~\citep{fisker2004}; the recurrence times from the two steps release of the thermonuclear energy model are two orders of magnitude larger than observations~\citep{1985ApJ...299..487S}; the observed multipeaked X-ray burst from 4U 1636-53 against burst induced accretion-disk corona model~\citep{1987ApJ...321L..67P}; the flame spreading model has drawbacks that triple-peaked and quadruple-peaked bursts cannot be explained~\citep{zhang2009,li2021}. Moreover, these models cannot reproduce the light curve of the observed double-peaked bursts~\citep{bhattacharyya2006A,li2021}.

\cite{ayasli1982} proposed a double-peaked structure using the stellar evolution code ASTRA, the first very sharp peak results from the helium burning, and the second peak results from the energy generation of mixed hydrogen/helium flashes via the $rp$ process. Such a double-peaked profile was also noticed by \cite{1984PASJ...36..199H}. Both works adopted the approximation network, \cite{ayasli1982} assumed a 15$\,$s characteristic timescale to synthesize nickel from sulfur, which leads to a broad peak separation time.  \cite{1984PASJ...36..199H} adopted 2.327$\,$s instead of 15$\,$s, which results in a very short peak separation time as 1.8$\,$s. Both failed to explain 4-7$\,$s peak separation of the double-peaked burst observations~\citep{li2021}.

\cite{fisker2004} suggested a nuclear waiting point impedance in the thermonuclear reaction flow to explain the double peak structure using the modified version of AGILE~\citep{2002ApJS..141..229L}. However, only the nuclear reaction waiting points: $^{22}\rm Mg$, $^{26}\rm Si$, $^{30}\rm S$, $^{34}\rm Ar$ are considered, the impact of the $rp$ process potential waiting points: $^{56}{\rm Ni}$, $^{60}\rm Zn$, $^{64}\rm{Ge}$, $^{68}\rm{Se}$, $^{72}\rm{Kr}$ on the double peak structure have not been explored~\citep{2006NuPhA.777..601S}. \cite{lampe2016} simulated the nuclear origins of double-peaked bursts using the KEPLER code, and they found that the double-peak structures are prominent in the higher metallicity ($Z=0.1$) and low accretion rates models. However, a larger than typical metallicity ($Z=0.01$) is required to produce the double-peaked light curves, which conflicts with the models in \cite{ayasli1982}, where the double-peaked burst is produced with $Z=0.01$. Recently, \cite{bult2019} interpreted that the bright double-peaked bursts are due to the local Eddington limits associated with the H and He layers of the NS envelope.

Understanding the above mechanisms is important because the light curves and temperature profiles of the double-peaked bursts, as well as triple-peaked/quadruple-peaked bursts, have not yet been successfully explained. A detailed survey of the available parameter space such as mass accretion rate and metallicity is important to understand the nuclear origins of the double-peaked bursts. Besides, the base luminosity which depends on $Q_{\rm b}$ and $\dot{M}$, has a significant impact on the typically single-peaked type I X-ray burst~\citep{2018ApJ...860..147M,2021ApJ...923...64D,2023ApJ...950..110Z} and the frequency of the mHz quasi-period oscillations~\citep{2014ApJ...787..101K,2023MNRAS.525.2054L}. As a result, the investigation of the impact of $Q_{\rm b}$ on the double-peaked structure is one of our purposes in this work.

In this study, we simulate the nuclear origins of the double-peaked type I X-ray bursts using the MESA (Modules for Experiments in Stellar Astrophysics) code~\citep{paxton2011,paxton2013,paxton2015,2018ApJS..234...34P}. We describe the details of the input physics, reaction network, and fuel composition in Section ~\ref{sec:model}. In Section~\ref{sec:inf}, we present our simulations of double-peaked bursts with high and ordinary metallicity, respectively. The impacts of mass accretion rate, metallicity, base heating, and more nuclear reaction waiting points on the double-peaked structure are discussed, and the parameter space for the conditions of double-peaked bursts is found. In Section~\ref{sec:obs}, we compare our results with the observations from 4U 1636-53 and 4U 1730-22. The composition of ashes from double-peaked bursts is also explored. We summarize our conclusions in Section~\ref{sec:con}.

\section{Model}\label{sec:model}
We utilize the one-dimensional stellar evolution code MESA version 9793 to simulate double-peaked bursts under different conditions. The details of the numerical approach and physics models can be found in the associated instrumentation papers~\citep{paxton2011,paxton2013,paxton2015,2018ApJS..234...34P}. Here, we summarize the most relevant details for this work. The envelope is 0.01{\,}km thick, with an inner boundary of NS mass $M=1.4\,M_{\odot}$ and radius $R=11.2\,\rm km$, which is the same as ~\citep{2018ApJ...860..147M}. We adopt base heat $Q_{\rm b}=0.1,0.2,0.3,0.4\,\rm MeV/u$ to consider the heat flow from the crust into the envelope, which is consistent with the constraint of $Q_{\rm b}<0.5\,\rm MeV/u$~\citep{2018ApJ...860..147M} and $Q_{\rm b}\simeq0.3-0.4\,\rm MeV/u$ calculated by ~\cite{2021ApJ...923...64D}. In MESA, this is achieved by fixing the luminosity at the base of the envelope, the base luminosity can be written as $L_{\rm base} = \dot{M}Q_{\rm b}$. The mass accretion rate is adopted the values close to $4.75\times10^{-10}\,M_{\odot}/\rm{yr}$ and $(0.2-1)\times10^{-8}\,M_{\odot}/\rm {yr}$ that are responsible for the double-peaked bursts in the previous works, the former is based on \cite{ayasli1982} and the latter is based on \cite{lampe2016}. The composition of the accreted fuel is determined by assuming the helium fraction ($Y$) changes with metallicity according to $Y=0.24+1.75Z$ ~\citep{lampe2016}, then the fraction of hydrogen ($X$) can be inferred by $X=1-Y-Z$.

The envelope is discretized into $\sim1000$ zones, and the local gravity in a zone is corrected for general relativity(GR) effects using a post-Newtonian correction, which is achieved by setting ``\textit{use\_GR\_factors = .true.}" ~\citep{paxton2011,paxton2015}. Adaptive time and spatial resolution were employed by setting ``\textit{varcontrol\_target = 1d-3}'' and ``\textit{mesh\_delt\_coeff = 1.0}"~\citep{2018ApJ...860..147M}.

We adopted the $rp.net$ nuclear reaction network, which includes 304 isotopes~\citep{fisker2007}, and the nuclear reaction rates from the REACLIB V2.2 library~\citep{cyburt2010}. 

The nuclear waiting point impedance is explored by variation of the $^{22}\rm Mg(\alpha,p)^{25}\rm Al$, $^{26}\rm{Si}(\alpha,p)^{29}\rm P$, $^{30}\rm S(\alpha,p)^{33}\rm Cl$, $^{34}\rm{Ar}(\alpha,p)^{37}\rm K$, $^{56}\rm{Ni}(p,\gamma)^{57}\rm Cu$, $^{60}\rm{Zn}(p,\gamma)^{61}\rm{Ga}$, $^{64}\rm{Ge}(p,\gamma)^{65}\rm {As}$, $^{68}\rm{Se}(p,\gamma)^{69}\rm {Br}$, $^{72}\rm{Kr}(p,\gamma)^{73}\rm{Rb}$ reaction up and down via a factor 100~\citep{fisker2004,cyburt2016}, where the factor roughly means the reactions had been underestimated (up) or overestimated (down) by a factor of 100 for the reaction rate uncertainty~\citep{2021PhRvL.127q2701H} .

A series of double-peaked bursts were simulated with models differed in mass accretion rate ($\dot{M}$), composition (X,Y,Z), base heat ($Q_{\rm b}$) and the nuclear reaction waiting points.

\section{results}\label{sec:inf}
Using MESA, we have carried out a series of numerical calculations to study the properties of double-peaked bursts. Our method and the physics inputs that we employed are described in the preceding section. The two different nuclear origins of double-peaked type I X-ray bursts are obtained (standard model 1 and standard model 2). The effects of mass accretion rate, metallicity, base heating, as well as the nuclear reaction waiting points on the properties of double-peaked structure are examined.

\subsection{ The nuclear origins of double-peaked type I X-ray bursts}
\cite{ayasli1982} studied the thermonuclear flashes on accreting neutron stars using a modified version of the stellar evolution code ASTRA, an approximation network including twenty nuclear species was used. The double-peaked structure was found with parameters: $\dot{M}=4.75\times10^{-10}\,M_{\odot}/\rm yr$, fuel composition $X=0.69$, $Y=0.30$ and $Z=0.01$. The explanation corresponding to the double peaks is as follows: the first peak results from helium burning, the depletion of helium causes the dip in the luminosity,
and the subsequent consumption of hydrogen via the $rp$ process leads to the second peak. \cite{lampe2016} simulated thermonuclear X-ray bursts by use of the KEPLER code, they found that the double-peaked structures are prominent in the higher metallicity models ($Z=0.1$) at low mass accretion rates ($\dot{M}=3.5\times10^{-9}\,M_{\odot}/{\rm yr}$), as mass accretion rate increases, the two peaks become indistinguishable. Their explanation is as follows: the first peak is a helium flash, which convects rapidly to the surface in less than a second, and the second peak occurs due to a nuclear waiting point impedance. 

\begin{figure}
    \centering
    \begin{minipage}{1.0\linewidth}
        \centering
        \includegraphics[width=\linewidth]{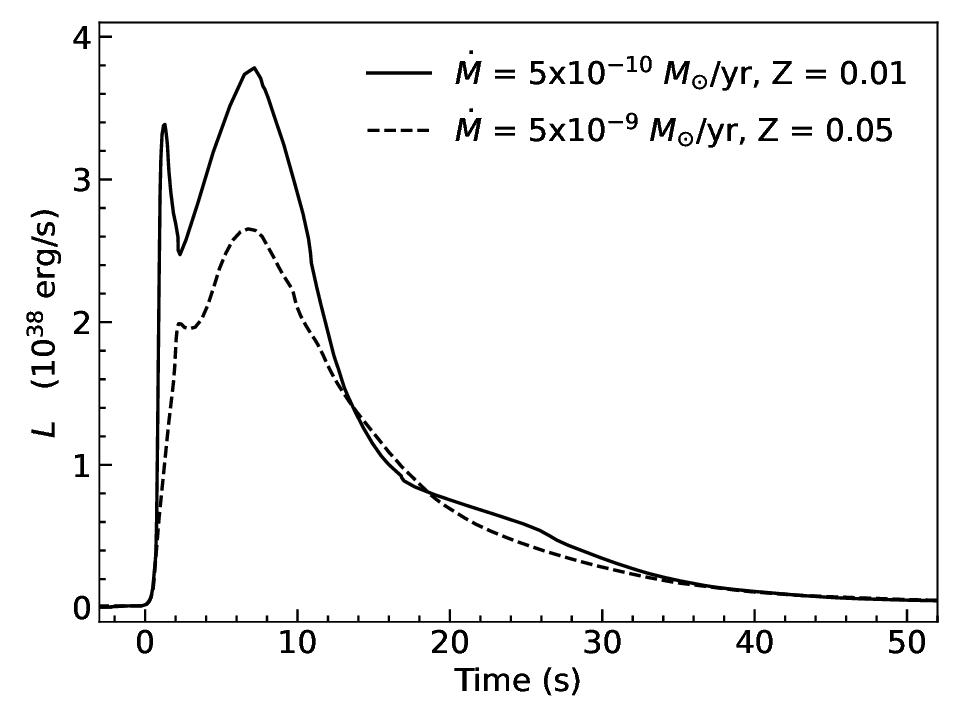}
    \end{minipage}
    \caption{Double-peaked bursts with low mass accretion rate($\dot{M}=5\times10^{-10}\,M_{\odot}/\rm yr$) at metallicity $Z=0.01$ (solid line) and an order of magnitude higher mass accretion rate $\dot{M}=5\times10^{-9}\,M_{\odot}/\rm yr$ at $Z=0.05$ (dashed line). For convenience, the former is called standard model 1, the latter is called standard model 2. In both models, the value of base heating is set as $Q_{\rm b}=0.1\,\rm MeV/u$.}
    \label{fig:m}
\end{figure}

Both the accretion rate and metallicity in ~\cite{ayasli1982} (for convenience, we call it case 1) are one order of magnitude lower than that in ~\cite{lampe2016}(case 2). So far, only a little attention has been paid to studying the conditions for double-peaked bursts. Here, we explore it with the use of MESA. By sequences of simulations with different mass accretion rates and metallicity, we obtained the double-peaked bursts corresponding to the above two scenarios. The results are shown in Fig.~\ref{fig:m}. The presence of double-peaked structure in our models with low mass accretion rate ($\dot{M}=5\times10^{-10}\,M_{\odot}/\rm yr$) at ordinary metallicity ($Z=0.01$) and an order of magnitude higher mass accretion rate ($\dot{M}=5\times10^{-9}\,M_{\odot}/\rm yr$) at $Z=0.05$ indicates that the nuclear origins of the double-peaked bursts can be simulated consistently with MESA. The rest physics inputs are the same between the two models, e.g., $M=1.4\,M_{\odot}$, $R=11.2\,\rm km$, $Q_{\rm b}=0.1\,\rm MeV/u$, which can be found in Table~\ref{tab:input} from the Appendix. These two models are marked as our standard models in two cases. We can find that the peak luminosity and peak ratio are different in the two cases. The dips between the two peaks are more obvious with lower mass accretion rate and metallicity (standard model 1). From the perspective of observations, the rise time of the first and second peaks, the separation time between the two peaks, as well as the peak flux ratio are varied among the samples of double-peaked bursts~\citep{li2021}. It's worth investigating whether double-peaked bursts can have a nuclear origin influenced by different factors, such as mass accretion rate, metallicity, base heating, and nuclear reaction waiting points. These factors will be explored in the following.  

\subsection{Impact of mass accretion rate}

\begin{figure}
    \centering
    \begin{minipage}{1.0\linewidth}
        \centering
        \includegraphics[width=\linewidth]{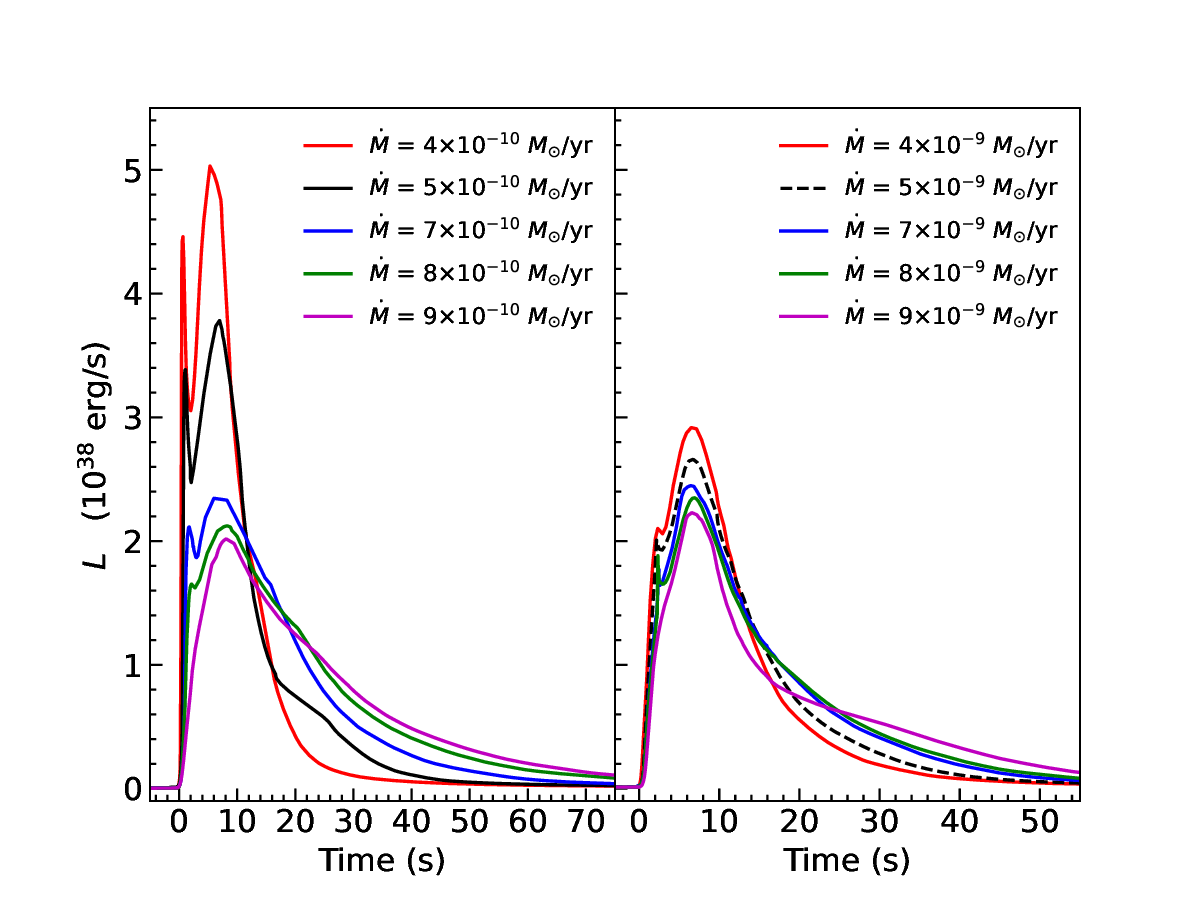}
    \end{minipage}
    \caption{
    Double-peaked bursts for the standard model and additional models with the same parameter values as those of the standard model except for the mass accretion rate. 
    Left: Light curves as functions of mass accretion rate based on standard model 1. Right: Light curves as a function of mass accretion rate based on standard model 2. In both cases, the double-peaked structures are prominent in the low mass accretion rate and will disappear as the mass accretion rate increases.}
    \label{fig:ml}
\end{figure}

\begin{figure}
    \centering
    \begin{minipage}{1.0\linewidth}
        \centering
        \includegraphics[width=\linewidth]{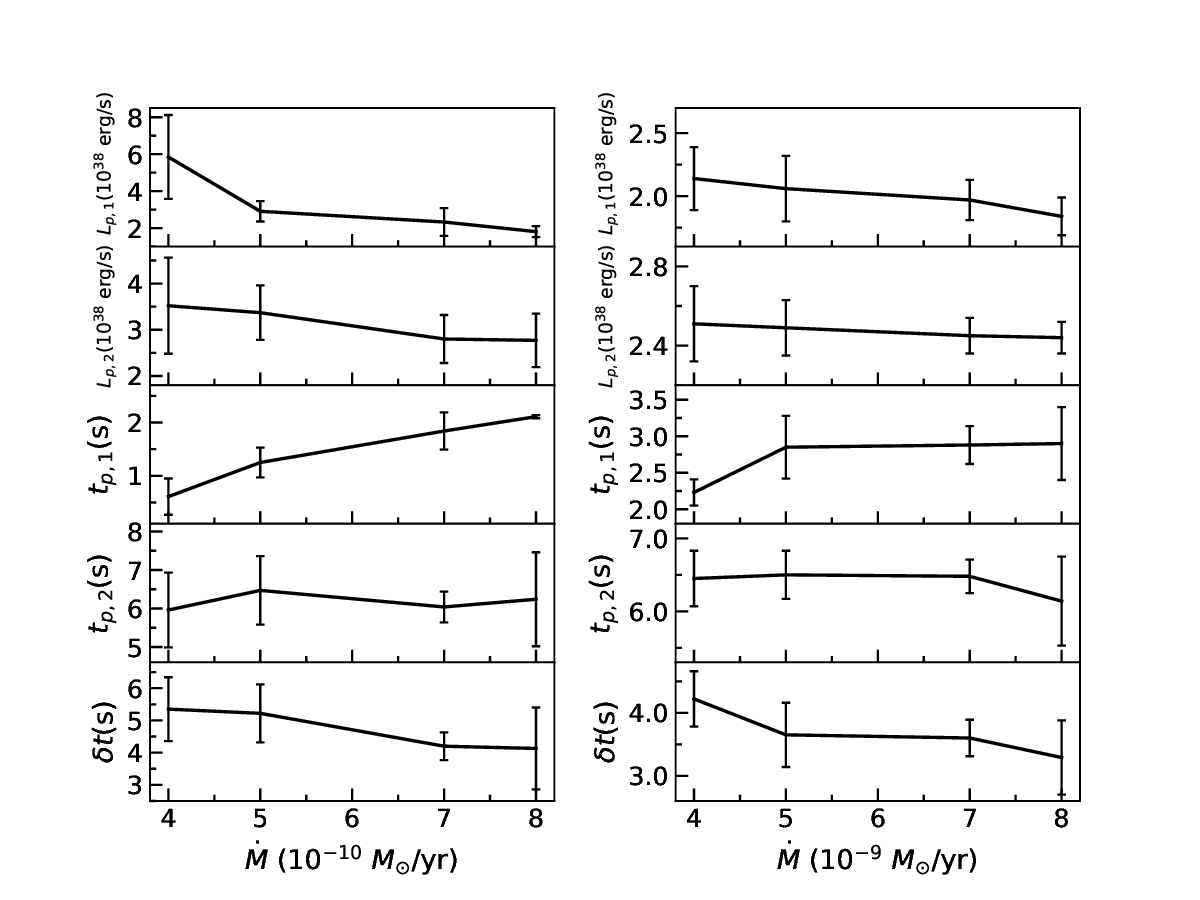}
    \end{minipage}
    \caption{
    The averaged values with $1\sigma$ errorbars of the first and second peak luminosity ($L_{\rm p,1}$, $L_{\rm p,2}$), the rise time of the first and second peak ($t_{\rm p,1}$, $t_{\rm p,2}$), and peak interval time $\delta t$ as a function of $\dot{M}$. Left: models 3-6 in case 1 (corresponding to the left panel of Fig.~\ref{fig:ml}). Right: models 7-10 in case 2 (corresponding to the right panel of Fig.~\ref{fig:ml}).
    }
    \label{fig:mt}
\end{figure}

Mass accretion rate plays an important role in type I X-ray burst~\citep{fujimoto1981,golloway2021}. At high mass accretion rate ($\dot{M}\gtrsim 10^{-8}\,M_{\odot}/\rm yr$), the accreted fuel burns stably. At relatively high mass accretion rate ($10^{-9}\,M_{\odot}/ {\rm yr} \lesssim \dot{M} \lesssim 10^{-8}\,M_{\odot}/\rm yr$), the mixed H/He burst occurs with a long tail from $rp$ process. At low mass accretion rate ($10^{-10}\,M_{\odot}/ {\rm yr} \lesssim \dot{M} \lesssim 10^{-9}\,M_{\odot}/\rm yr$), pure helium X-ray burst occurs. At still lower mass accretion rate ($\dot{M}\lesssim 10^{-10}\,M_{\odot}/\rm yr$), as the unstable hydrogen burning occurs before the unstable helium burning, the mixed H/He burst occurs. There have been many efforts to explore the effects of mass accretion rate on thermonuclear bursts~\citep{ayasli1982,lampe2016,2018ApJ...860..147M,2020MNRAS.494.4576J}, which leads us to understand more clearly about the $\dot{M}$ dependent type I X-ray burst. However, the parameter space of $\dot{M}$ for double-peaked bursts is unclear. We varied the mass accretion rate based on standard models 1 and 2. The results are illustrated in Fig.~\ref{fig:ml}.

We have plotted the 11th burst for each simulation in Fig.~\ref{fig:ml} to ensure a consistent ashes layer following~\cite{lampe2016}. The double-peaked structures are prominent in the lower mass accretion rate models, while the helium peak is suppressed at higher mass accretion rate in the two cases, which is consistent with ~\cite{lampe2016}. In case 1 (left panel of Fig.~\ref{fig:ml}), the double-peaked structure appears at mass accretion rate in the range of $\sim(4-8)\times10^{-10}\,M_{\odot}/\rm yr$ for $Z=0.01$. The double-peaked structure disappears at $\dot{M}=9\times10^{-10}\,M_{\odot}/\rm yr$ for $Z=0.01$. In case 2 (right panel of Fig.~\ref{fig:ml}), the double-peaked structure appears at mass accretion rate in the range of $\sim(4-8)\times10^{-9}\,M_{\odot}/\rm yr$ for $Z=0.05$. The double-peaked structure disappears at $\dot{M}=9\times10^{-9}\,M_{\odot}/\rm yr$ for $Z=0.05$.

The results for the parameter variation with mass accretion rate are shown in Fig.~\ref{fig:mt}.  With increase of $\dot{M}$, both the first and second peak luminosities($L_{\rm p,1}$, $L_{\rm p,2}$) decrease, the first peak time ($t_{\rm p,1}$) increases, the second peak time ($t_{\rm p,2}$) is almost constant, the peak separation time $\delta t$ decreases. This is because at a low mass accretion rate, the first peak is a helium flash with a very short rise time, with the increase of the mass accretion rate, the amount of pure helium layer at the base of the accreted material decreases, and mixed H/He burst occurs, as a result, the rise time of the first peak increases, the peak separation time decreases and the peak luminosity decreases. The changes in the two cases are almost consistent. The values of $L_{\rm p,1}$, $L_{\rm p,2}$, $t_{\rm p,1}$, $t_{\rm p,2}$, and $\delta t$ can be found in Table~\ref{tab:output} from the Appendix. Besides, the other properties for double-peaked bursts such as burst strength ($\alpha$), burst energy ($E_{\rm burst}$), recurrence time ($\Delta{\rm T}$), as well as the first ($L_{\rm p,1}$) and second ($L_{p,2}$) peak luminosity are also obtained and illustrated in Table~\ref{tab:output} from the Appendix. We discard the first two bursts and the last burst of each model to calculate these values with $1\sigma$ uncertainty. The number of bursts and bursts with two peaks of each model are shown in Table 1 in the appendix.

\subsection{Impact of metallicity}

\begin{figure}
    \centering
    \begin{minipage}{1.0\linewidth}
    \centering
         \includegraphics[width=\linewidth]{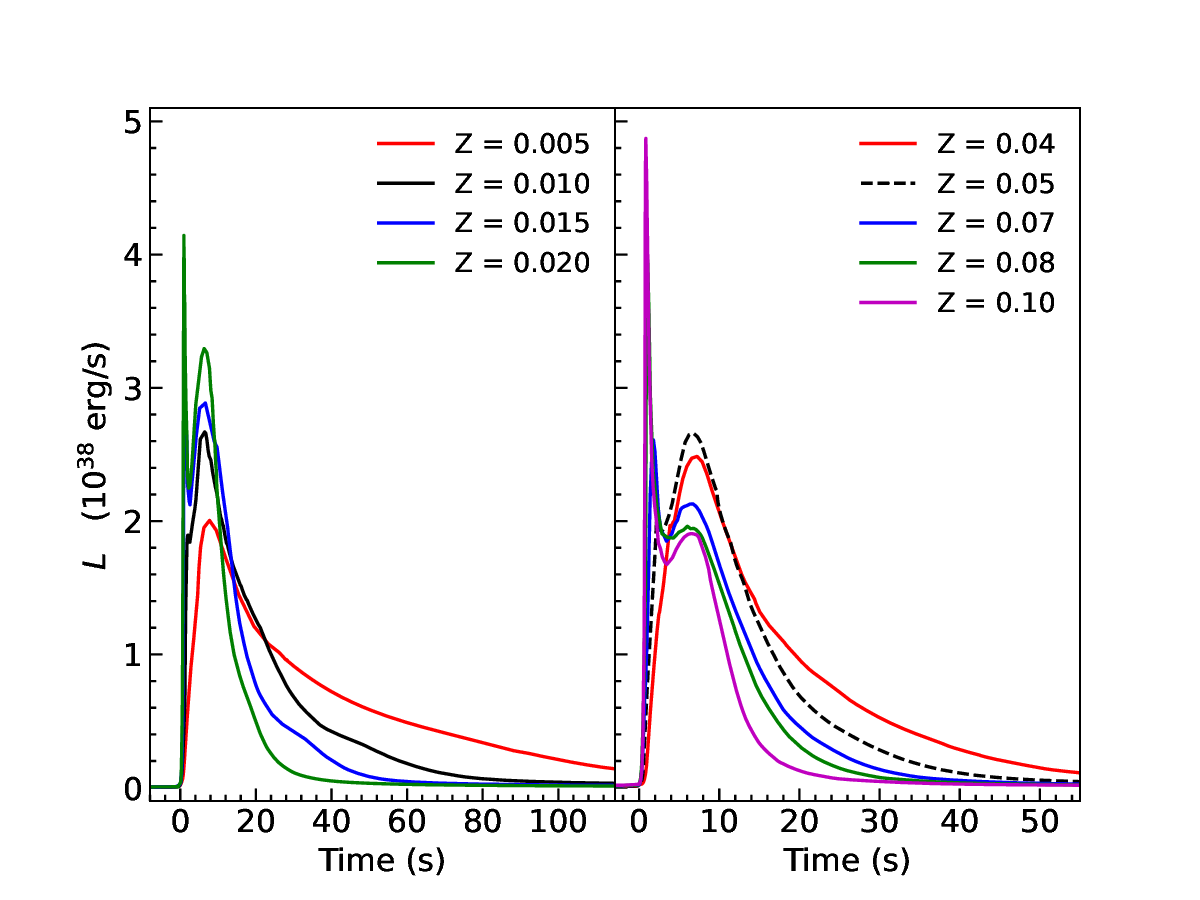}
         \end{minipage}    
    \caption{
    Same as Fig.~\ref{fig:ml} for the standard model and additional models with the same parameter values as those of the standard model except for the fuel composition due to the change of metallicity. 
    Left: Light curves as a function of metallicity based on model 5. Right: Light curves as a function of metallicity based on standard model 2. In both cases, the double-peaked structures are prominent in the relatively higher metallicity and will disappear as metallicity decreases.}
    \label{fig:zl}
\end{figure}

\begin{figure}
    \centering
    \begin{minipage}{1.0\linewidth}
    \centering
         \includegraphics[width=\linewidth]{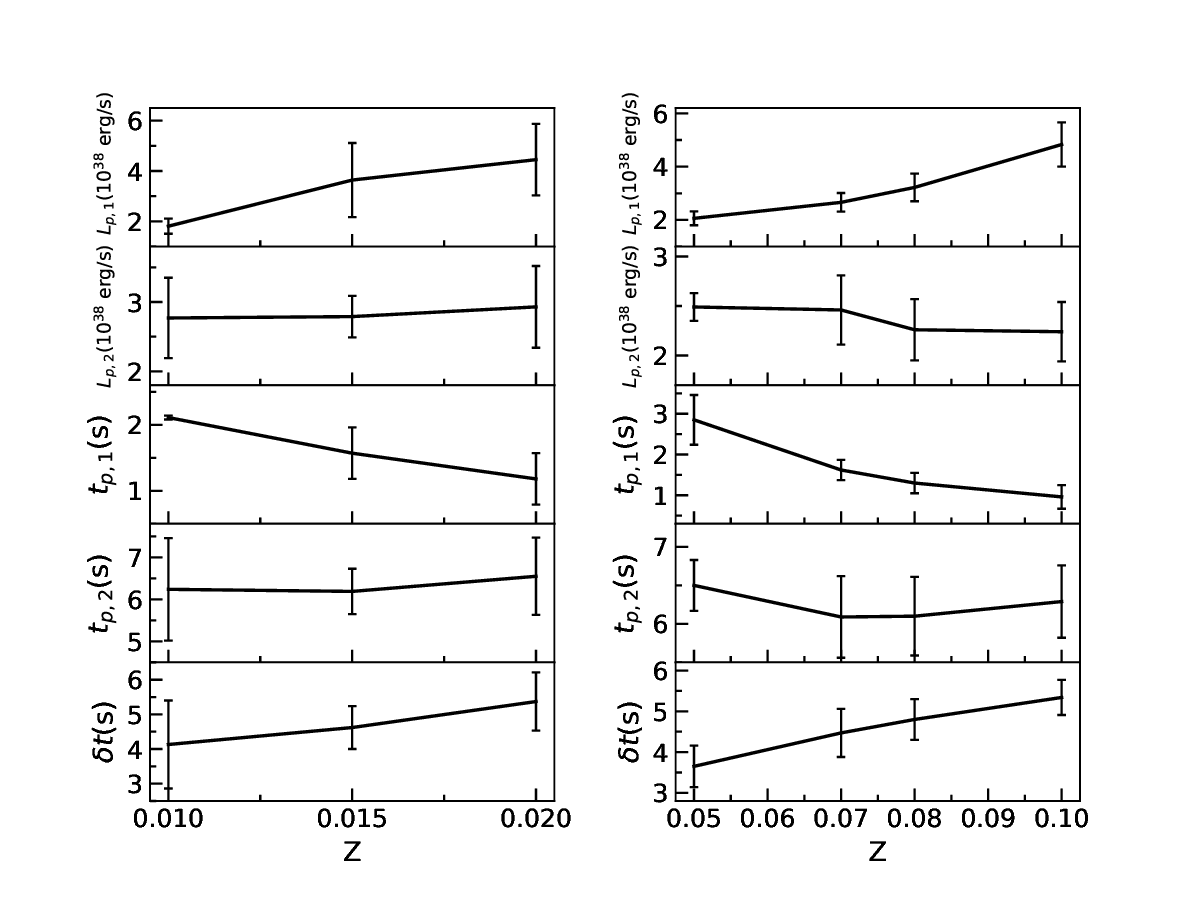}
         \end{minipage}    
    \caption{
    Same as Fig.~\ref{fig:mt} but as a function of $Z$. Left: models 11-13 in case 1 (corresponding to the left panel of Fig.~\ref{fig:zl}). Right: models 14-17 in case 2 (corresponding to the right panel of Fig.~\ref{fig:zl}).
    }
    \label{fig:zt}
\end{figure}

The metallicity of the accreting matter has an important influence on the properties of thermonuclear bursts. We present the results for several models wherein the initial metallicity of accreting matter is varied in their standard values. The results are shown in Fig.~\ref{fig:zl}. We find that the double-peaked structures are prominent in the high metallicity models, the first helium peak will be suppressed at lower metallicity in the two cases.

Fig.~\ref{fig:zt} shows the parameter variation with metallicity. The first peak luminosity $L_{\rm p,1}$ increases as metallicity increases, while the second peak luminosity $L_{\rm p,2}$ changes irregularly as $Z$ increases. The first peak time $t_{\rm p,1}$ decreases as $Z$ increases, and the second peak time $t_{\rm p,2}$ almost remains constant as $Z$ increases, resulting in the peak separation time increases. 

As the metallicity catalyzes the burning of hydrogen via the CNO cycle, increased $z$ increases the pure helium layer at the base of the accreted matter. An increased pure helium layer causes the first peak luminosity to increase and the first peak to rise rapidly.

\subsection{Impact of base heating}

\begin{figure}
    \centering
    \begin{minipage}{1.0\linewidth}
    \centering
         \includegraphics[width=\linewidth]{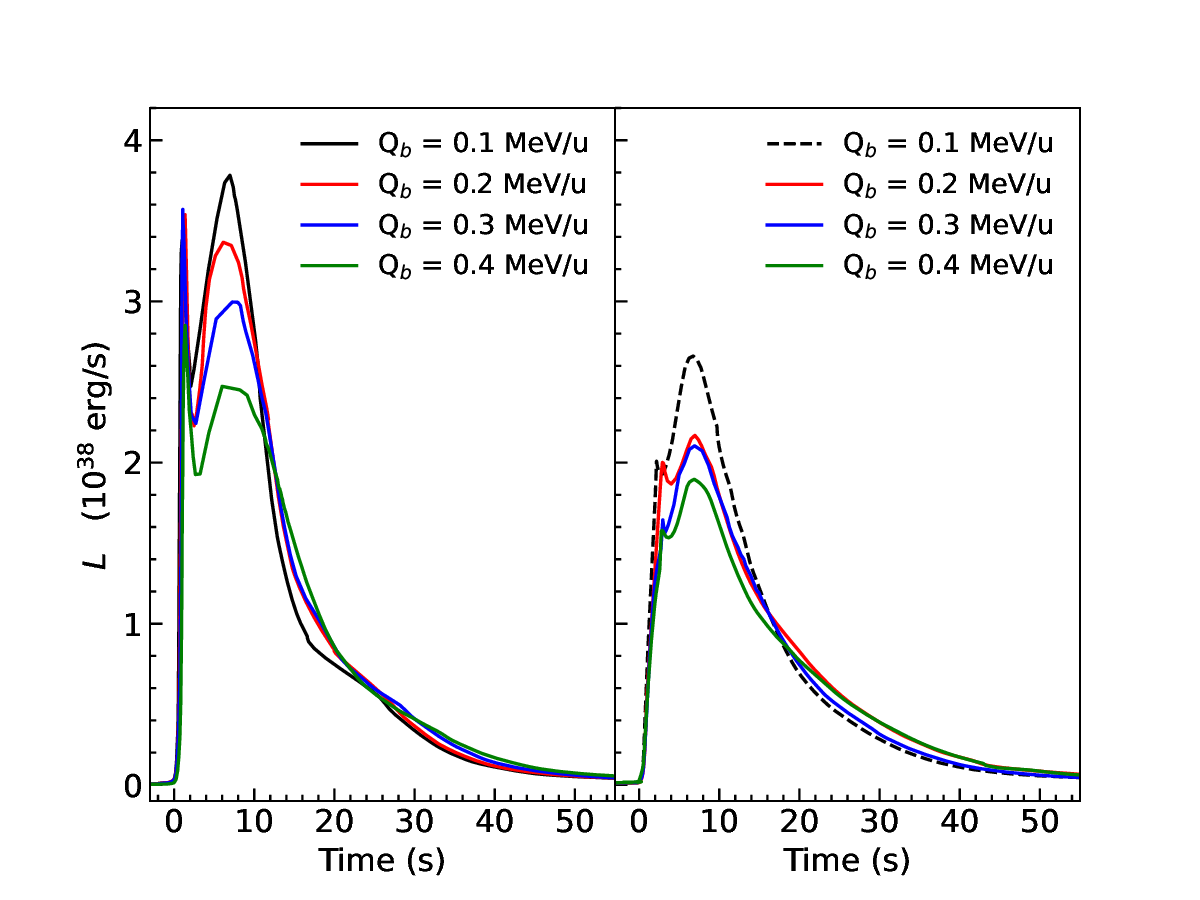}
         \end{minipage} 
    \caption{
    Same as Fig.~\ref{fig:ml} for the standard model and additional models with the same parameter values as those of the standard model except for the base heating $Q_{\rm b}$. 
    Left: Light curves as a function of base heating based on standard model 1. Right: Light curves as a function of base heating based on standard model 2. In both cases, the double-peaked structures will not disappear as $Q_{\rm b}$ increases.
    } 
    \label{fig:ql}
\end{figure}

\begin{figure}
    \centering
    \begin{minipage}{1.0\linewidth}
    \centering
         \includegraphics[width=\linewidth]{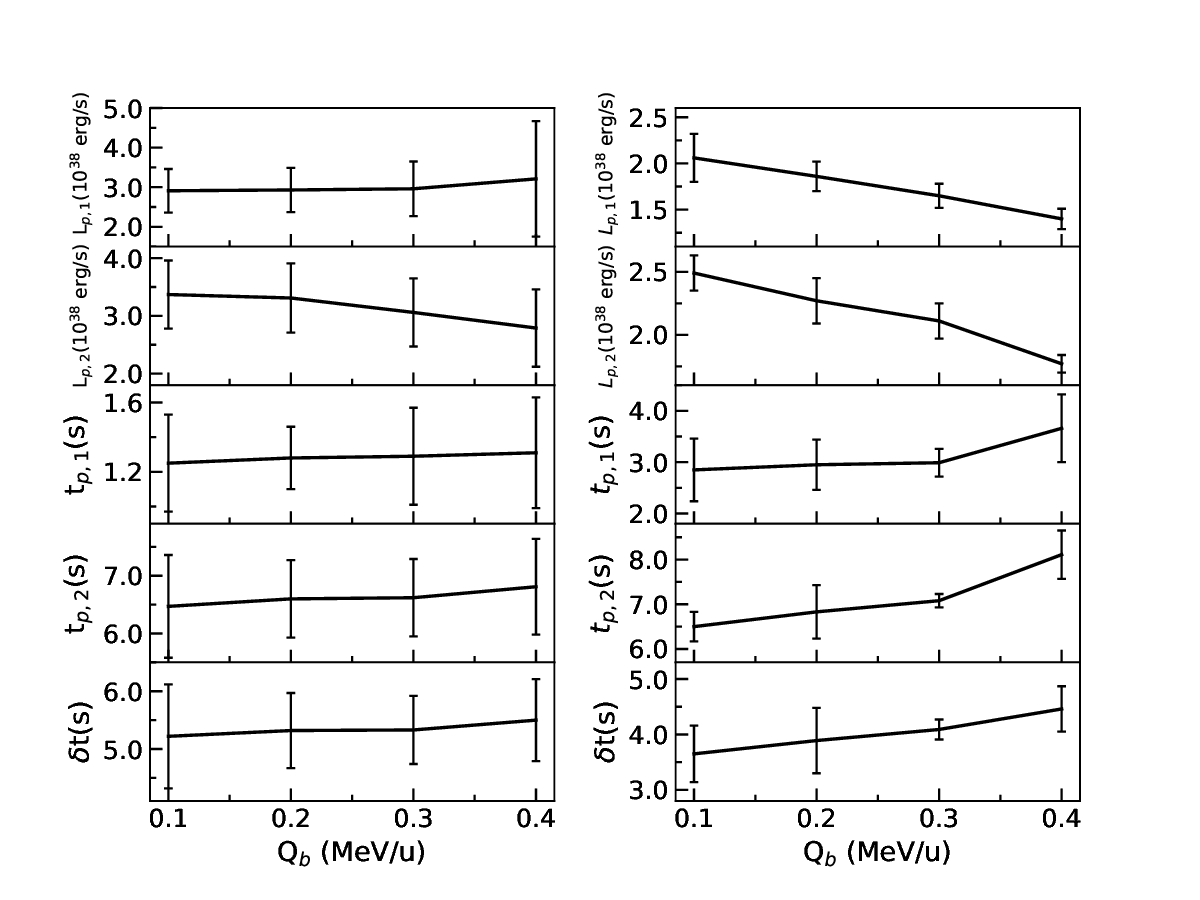}
         \end{minipage} 
    \caption{
    Same as Fig.~\ref{fig:mt} but as a function of $Q_{\rm b}$. Left: models 18-20 in case 1 (corresponding to the left panel of Fig.~\ref{fig:ql}). Right: models 21-23 in case 2 (corresponding to the right panel of Fig.~\ref{fig:ql}).
    }
    \label{fig:qt}
\end{figure}

The base heating is included for type I X-ray burst simulations which consider the NS envelope~\citep{2007ApJ...671L.141H,2018ApJ...860..147M,2020MNRAS.494.4576J}, this additional heating represents the heat flowing from the underlying NS crust to the envelope, and specified by the parameter $Q_{\rm b}$. Most models adopt a value of $Q_{\rm b}=0.1-0.15\,\rm MeV/u$~\citep{1990A&A...227..431H,golloway2021}. \cite{2018ApJ...860..147M} obtained the constraint of $Q_{\rm b}<0.5\,\rm MeV/u$ from the observations of GS 1826-24, \cite{2021ApJ...923...64D} calculated the values of $Q_{\rm b}\simeq0.3-0.4\,\rm MeV/u$ with use of HERES code, which covers the entire NS regions. We explore the effects of $Q_{\rm b}$ on the properties of double-peaked bursts. The results are presented in Fig.~\ref{fig:ql}. We find that the double-peaked structures are prominent at lower base heating and will not disappear as $Q_{\rm b}$ increases. 

Fig.~\ref{fig:qt} exhibits the parameter variation with base heating. We can see from the left panel of Fig.~\ref{fig:qt} that the first peak luminosity $L_{\rm p,1}$, the first and second peak time, as well as the peak separation time, remain nearly constant, the second peak luminosity $L_{\rm p,2}$ decreases as $Q_{\rm b}$ increases. In the right panel of Fig.~\ref{fig:qt}, we can see that both the first and second peak luminosities decrease as $Q_{\rm b}$ increases, the first peak rise time $t_{\rm p,1}$, the second peak time $t_{\rm p,2}$ and the peak separation time $\delta t$ increase as $Q_{\rm b}$ increases. This is because the hydrogen burning via hot CNO cycle lasts longer with smaller $Q_{\rm b}$, as a result, as $Q_{\rm b}$ increases, the amount of pure helium layer decreases, which leads to the decrease of the peak luminosity and increase of the peak time. Our results indicate that the $Q_{\rm b}$ effect is more sensitive to the double-peaked bursts in case 2.

\subsection{Impact of the nuclear reaction waiting points}

\begin{table}
    \centering
    \caption{
    The nuclear reaction waiting points impact the double-peaked bursts in the standard model 1. 
    }
    \begin{threeparttable}
    \begin{tabular}{cccc}
    \toprule
    Rank & Reaction & Type$^a$ & Sensitivity$^{b}$ \\
    \hline
    1 & $\rm ^{68}Se(p,\gamma) ^{69}Br$ & U & 14.93 \\
    2 & $\rm ^{22}Mg(\alpha,p) ^{25}Al$ & D & 14.71 \\
    3 & $\rm ^{60}Zn(p,\gamma) ^{61}Ga$ & U & 13.54 \\
    4 & $\rm ^{26}Si(\alpha,p) ^{29}P$ & D & 12.71 \\
    5 & $\rm ^{22}Mg(\alpha,p) ^{25}Al$ & U & 10.32 \\
    6 & $\rm ^{60}Zn(p,\gamma) ^{61}Ga$ & D & 9.09 \\
    7 & $\rm ^{26}Si(\alpha,p) ^{29}P$ & U & 8.65 \\
    8 & $\rm ^{34}Ar(\alpha,p) ^{37}K$ & U & 7.71 \\
    9 & $\rm ^{30}S(\alpha,p) ^{33}Cl$ & U & 6.44 \\
    10 & $\rm ^{68}Se(p,\gamma) ^{69}Br$ & D & 5.56 \\
    11 & $\rm ^{30}S(\alpha,p) ^{33}Cl$ & D & 5.51 \\
    12 & $\rm ^{56}Ni(p,\gamma) ^{57}Cu$ & U & 4.61 \\
    13 & $\rm ^{56}Ni(p,\gamma) ^{57}Cu$ & D & 3.10 \\
    14 & $\rm ^{72}Kr(p,\gamma) ^{73}Rb$ & D & 2.71 \\
    15 & $\rm ^{64}Ge(p,\gamma) ^{65}As$ & D & 2.55 \\
    16 & $\rm ^{64}Ge(p,\gamma) ^{65}As$ & U & 2.53 \\
    17 & $\rm ^{72}Kr(p,\gamma) ^{73}Rb$ & U & 1.29 \\ 
    18 & $\rm ^{34}Ar(\alpha,p) ^{37}K$ & D & 0.97 \\
     \hline
    \end{tabular}
    \begin{tablenotes}
    \footnotesize
        \item[a] Up(U) and down(D) represent the reaction rate changes of $\times100$ and $/100$, respectively. 
        \item[b] $E_{LC}^{(k)}$ in units of $\rm 10^{37}$ erg.
    \end{tablenotes}
    \end{threeparttable}
    \label{tab:nu1}
\end{table}

\begin{figure}
    \centering
    \begin{minipage}{1.0\linewidth}
    \centering
         \includegraphics[width=\linewidth]{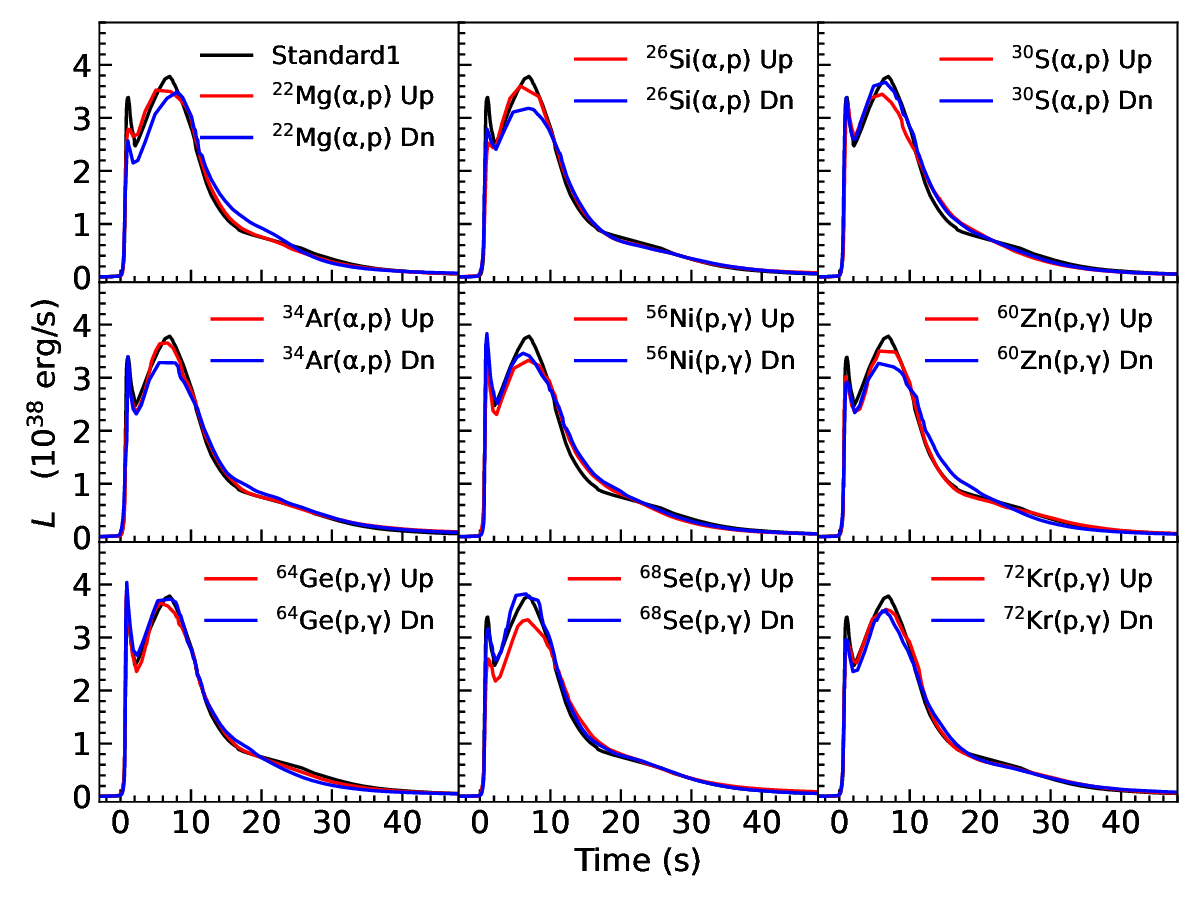}
         \end{minipage} 
    \caption{
    Changes in double-peaked burst light curves induced by variation of the nuclear reaction waiting points up (red) and down (blue), the dotted line represents the standard model 1. Up indicates a rate increase by a factor of 100, and Dn indicates a rate divided by a factor of 100.
    } 
    \label{fig:nb}
\end{figure}

\begin{table}
    \centering
    \caption{
    The nuclear reaction waiting points that impact the double-peaked bursts in the standard model 2.
    }
    \begin{threeparttable}
    \begin{tabular}{cccc}
    \toprule
    Rank & Reaction & Type$^a$ & Sensitivity$^{b}$ \\
    \hline
    1 & $\rm ^{68}Se(p,\gamma) ^{69}Br$ & D & 22.47 \\
    2 & $\rm ^{56}Ni(p,\gamma) ^{57}Cu$ & D & 11.36 \\
    3 & $\rm ^{30}S(\alpha,p) ^{33}Cl$ & U & 9.58 \\
    4 & $\rm ^{22}Mg(\alpha,p) ^{25}Al$ & D & 9.21 \\
    5 & $\rm ^{64}Ge(p,\gamma) ^{65}As$ & D & 8.71 \\
    6 & $\rm ^{22}Mg(\alpha,p) ^{25}Al$ & U & 8.46 \\
    7 & $\rm ^{34}Ar(\alpha,p) ^{37}K$ & U & 6.35 \\
    8 & $\rm ^{64}Ge(p,\gamma) ^{65}As$ & U & 6.33 \\
    9 & $\rm ^{60}Zn(p,\gamma) ^{61}Ga$ & U & 5.14 \\
    10 & $\rm ^{34}Ar(\alpha,p) ^{37}K$ & D & 4.73 \\
    11 & $\rm ^{68}Se(p,\gamma) ^{69}Br$ & U & 4.35 \\
    12 & $\rm ^{26}Si(\alpha,p) ^{29}P$ & U & 4.19 \\
    13 & $\rm ^{26}Si(\alpha,p) ^{29}P$ & D & 3.55 \\
    14 & $\rm ^{72}Kr(p,\gamma) ^{73}Rb$ & D & 2.85 \\
    15 & $\rm ^{30}S(\alpha,p) ^{33}Cl$ & D & 2.81 \\
    16 & $\rm ^{56}Ni(p,\gamma) ^{57}Cu$ & U & 1.89 \\ 
    17 & $\rm ^{60}Zn(p,\gamma) ^{61}Ga$ & D & 0.71 \\
    18 & $\rm ^{72}Kr(p,\gamma) ^{73}Rb$ & U & 0.29 \\
     \hline
    \end{tabular}
    \begin{tablenotes}
    \footnotesize
        \item[a] Up(U) and down(D) represent the reaction rate changes of $\times100$ and $/100$, respectively. 
        \item[b] $E_{LC}^{(k)}$ in units of $\rm 10^{37}$ erg.
    \end{tablenotes}
    \end{threeparttable}
    \label{tab:nu2}
\end{table}

\begin{figure}
    \centering
    \begin{minipage}{1.0\linewidth}
    \centering
         \includegraphics[width=\linewidth]{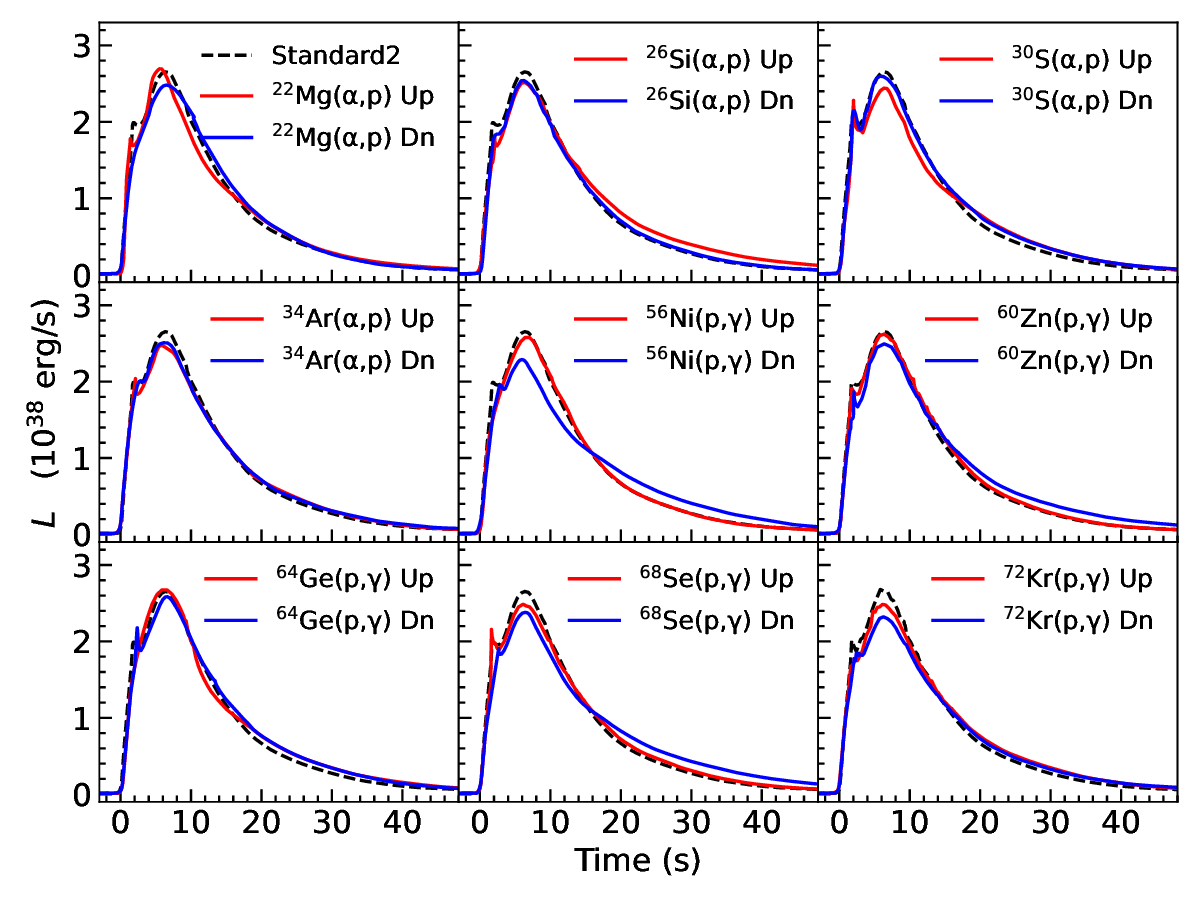}
         \end{minipage}   
    \caption{
    Same as Fig.~\ref{fig:nb} but the nuclear reactions are varied about the values used in standard model 2.
    }
    \label{fig:nd}
\end{figure}

\cite{fisker2004} proposed a nuclear waiting point impedance in the thermonuclear reaction flow to explain the double-peaked burst observations. The uncertainty in the theoretical $(\alpha,p)$-rates  $^{30}\rm S$ and $^{34}\rm Ar$ is studied. So far, no experimental data exist for the specific cases of $\alpha-$capture on $^{22}{\rm Mg} $, $^{26}{\rm Si}$, $^{30}{\rm S}$, $^{34}{\rm Ar}$ or proton-capture on $^{56}{\rm Ni}$, $^{60}{\rm Zn}$, $^{64}{\rm Ge}$, $^{68}{\rm Se}$ and $^{72}{\rm Kr}$, although~\cite{2021PhRvL.127q2701H} explored the new $^{22}{\rm Mg}(\alpha,p)^{25}{\rm Al}$ reaction rate without direct experiment measurement and found 6 orders of magnitude lower than the previous theoretical model. We systematically investigate the uncertainty in the theoretical $(\alpha,p)$-rates $^{22}\rm Mg$, $^{26}\rm Si$, $^{30}\rm S$, $^{34}\rm Ar$ and $(p,\gamma)$-rates $^{56}{\rm Ni}$, $^{60}\rm Zn$, $^{64}\rm{Ge}$, $^{68}\rm{Se}$, $^{72}\rm{Kr}$ on the double-peaked bursts. To explore this effect, we span the reaction rate change of $/100$ (down) to $\times100$ (up) following~\cite{cyburt2016}. Factor 100 was chosen to ensure no sensitivity is missed, and such large uncertainties were found in the theoretical reaction rates via the Hauser-Feshbach approach ~\citep{2000ADNDT..75....1R}.  

To quantify the influence of the reaction rate variation on double-peaked bursts, we adopt the sensitive value $E_{LC}^{k}$ defined in ~\cite{cyburt2016}:
\begin{eqnarray}
E_{LC} ^{(k)} &=& \int\mid\langle L_{k}(t)\rangle-\langle L_{0}(t)\rangle\mid dt
\end{eqnarray}
where $L_{k}(t)$ is the light curve of each variation $k$, $L_0(t)$ is the luminosity of the standard model.

Fig.~\ref{fig:nb} shows the changes in double-peaked burst light curves induced by variation in ($\alpha,p$) and ($p,\gamma$) reaction rates based on standard model 1. 
It is found that $^{68}\rm Se(p,\gamma)^{69}\rm{Br}$-reaction with a factor by $\times 100$ will decrease the luminosity of the two peaks, it is the most sensitive reaction on the double-peaked bursts. The sequence of the sensitivity of the nuclear reaction waiting points can be found in Table~\ref{tab:nu1}. 

Similarly, Fig.~\ref{fig:nd} shows the changes in double-peaked burst light curves induced by variation in ($\alpha,p$) and ($p,\gamma$) reaction rates but the nuclear reactions are varied about the values used in standard model 2. We find that the double-peaked structure will disappear if the $^{56}{\rm Ni}(p,\gamma)^{57}\rm Cu$-reaction and $^{64}{\rm Ge}(p,\gamma)^{65}\rm As$-reaction were faster by a factor 100, the $^{22}{\rm Mg}(\alpha,p)^{25}\rm Al$-reaction were slower by a factor 100. Table~\ref{tab:nu2} summarizes the nuclear reaction waiting points that impact the double-peaked light curve in a sequence of sensitivity.
In this case, $^{68}\rm Se(p,\gamma)^{69}\rm{Br}$-reaction with the reaction rate changes of /100 is the most sensitive reaction on the double-peaked bursts.

From the above analysis, we can find that the effect of the uncertainty in the theoretical reaction rates at waiting points is differed by the conditions of the double-peaked bursts. $^{22}\rm Mg$, $^{56}{\rm Ni}$, $^{64}{\rm Ge}$, $^{68}{\rm Se}$ are possibly the most important nuclear waiting points impedance in the thermonuclear reaction flow to explain the double-peaked bursts.

\section{Discussion}\label{sec:obs}
So far, only a little attention has been paid to the model-observation comparisons and the final products of the double-peaked bursts, we will discuss these issues with the above calculations.
\subsection{Model-observation comparisons}
\begin{table*}
    \centering
    \caption{
    Observational values of the peak ratio ($r_{1,2}$), the first ($t_{{\rm p},1}$) and second ($t_{{\rm p},2}$) peak time, and the peak separation time ($\delta t$) from 4U 1636-53~\citep{li2021} and 4U 1730-22~\citep{chen2023}.
    }
    \begin{tabular}{cccccccc}
    \toprule
    Source & mission & date & Obsid & $\rm r_{1,2}$ & $\rm t_{p,1}(s)$ & $\rm t_{p,2}(s)$ & $\delta t(\rm s)$ \\
    \hline
    4U 1636-53 & RXTE & 20011003 & 60032-01-13-01 & 0.73 & 3.10 & 7.10 & 4.00 \\
    4U 1730-22 & Insight-HXMT & 20220509 & P051400201003-20220509-01-01 & 0.76 & 1.00 & 8.50 & 7.50 \\
     \hline
    \end{tabular}
    \label{tab:gc1}
\end{table*} 

\cite{li2021} analysed 16 multipeaked type I X-ray bursts from 4U 1636-53 with the Rossi X-ray Timing Explorer(RXTE), where 14 double-peaked bursts were included. However, only one Obsid has been studied for its spectral evolution (as shown in Fig.$\,5$ from~\citep{li2021}). \cite{chen2023} studied 10 thermonuclear bursts from 4U 1730-22 with Insight-HXMT during its two outbursts in 2021 \& 2022, and found one double-peaked burst. We take the values of the first ($t_{{\rm p},1}$) and second ($t_{{\rm p},2}$) peak time, the first and second peak flux ratio ($r_{1,2}$) and the peak separation time ($\delta t$) from the spectral evolution of the bolometric flux about the above two observations, the values are illustrated in Table~\ref{tab:gc1}. 

\begin{figure}
    \centering
    \begin{minipage}{1.0\linewidth}
    \centering
         \includegraphics[width=\linewidth]{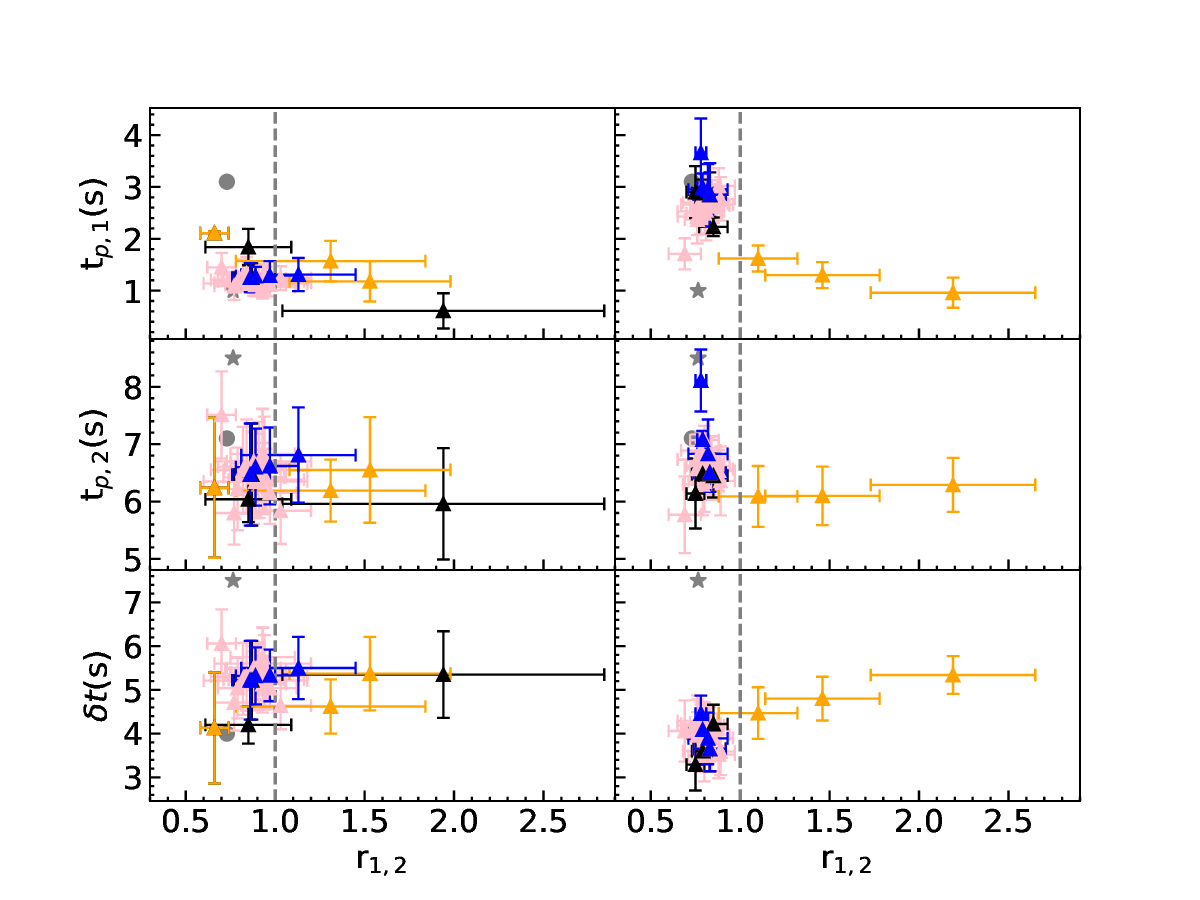}
         \end{minipage} 
         
    \caption{
    The first ($t_{{\rm p},1}$) and second ($t_{{\rm p},2}$) peak time, the peak separation time ($\delta t$) as a function of peak ratio ($r_{1,2}$). Black: the averaged values with $1\sigma$ errorbars under the variation in mass accretion rate. Orange: the averaged values with $1\sigma$ errorbars under the variation in metallicity. Blue: the averaged values with $1\sigma$ errorbars under the variation in base heating. Pink: the averaged values with $1\sigma$ errorbars under the variation in nuclear reaction waiting points. Solid circle: the values from the observations of 4U 1636-53. Star: the values from the observations of 4U 1730-22.
    }
    \label{fig:ob}
\end{figure}

Fig.~\ref{fig:ob} shows the comparison between our models and the observations shown in Table~\ref{tab:gc1}, where our theoretical models are varied in mass accretion rate, metallicity, base heating, as well as the uncertainty of the nuclear reaction waiting points. We can see that the values of the first peak time $t_{{\rm p},1}$ are in the range of $\sim 1-4\,\rm s$, the second peak time $t_{{\rm p},2}$ is in the range of $\sim 6-8\,\rm s$, the peak separation time is in the range of $\sim 3-6\,\rm s$, and the peak flux ratio is in the range of $\sim 0.5-2.5$. The observational values of 4U 1636-53 with Obsid 60032-01-13-01 can be well explained with the changes in the parameters based on standard model 2. However, our nuclear origins models still can not explain the double-peaked burst with the longer peak separation time (e.g. 4U 1730-22 with $\delta t>7\,\rm s$).

\subsection{Final products during double-peaked bursts nucleosynthesis}
\begin{figure}
    \centering
    \begin{minipage}{\linewidth}
    \centering
         \includegraphics[width=\linewidth]{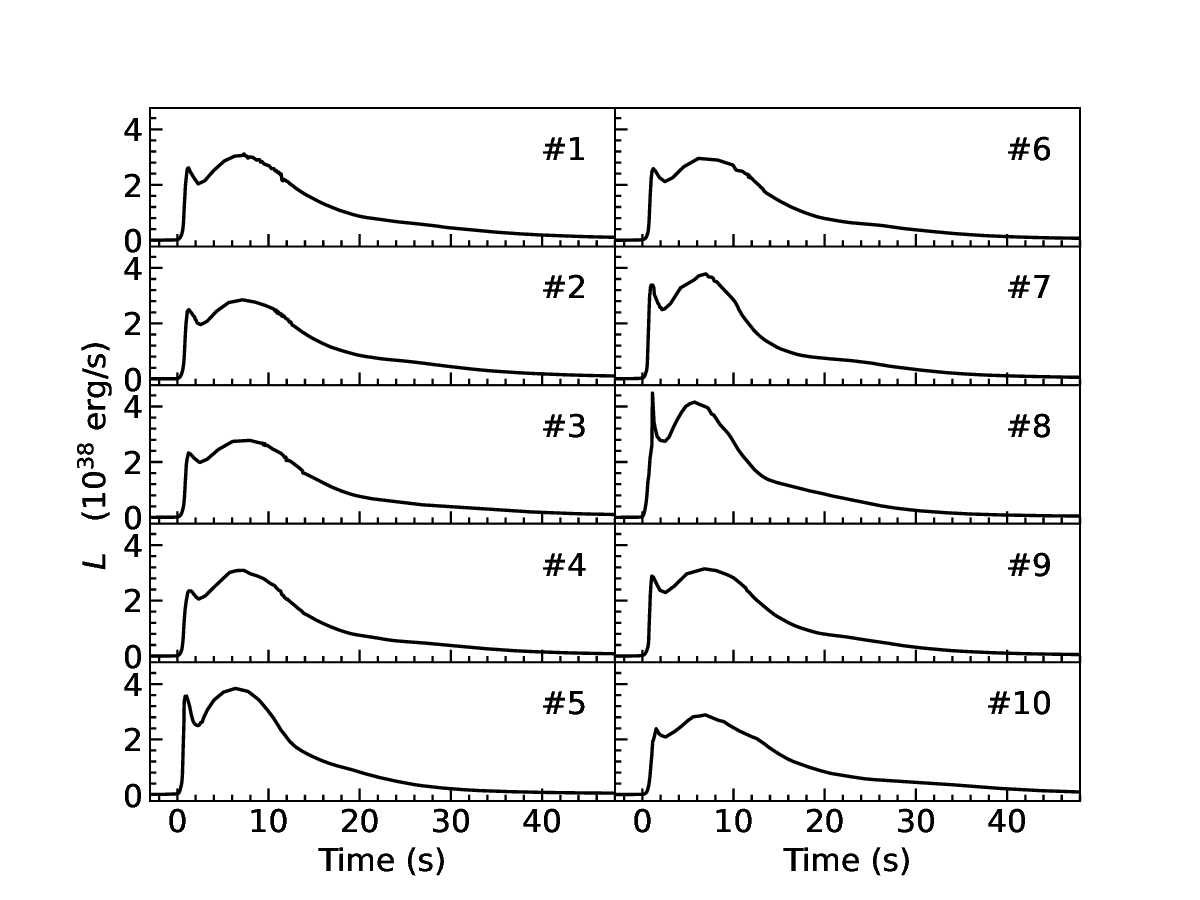}
         \end{minipage} 
         
    \caption{
    Samples of a succession of the double-peaked bursts for standard model 1 with $X=0.7325$, $Y=0.2575$, $Z=0.01$, $\dot{M}=5\times10^{-10}\,M_{\odot}/\rm yr$, $Q_{\rm b}=0.1\,\rm MeV/u$.}
    
    \label{fig:ash1}
\end{figure}

\begin{figure}
    \centering
    \begin{minipage}{1.0\linewidth}
    \centering
         \includegraphics[width=\linewidth]{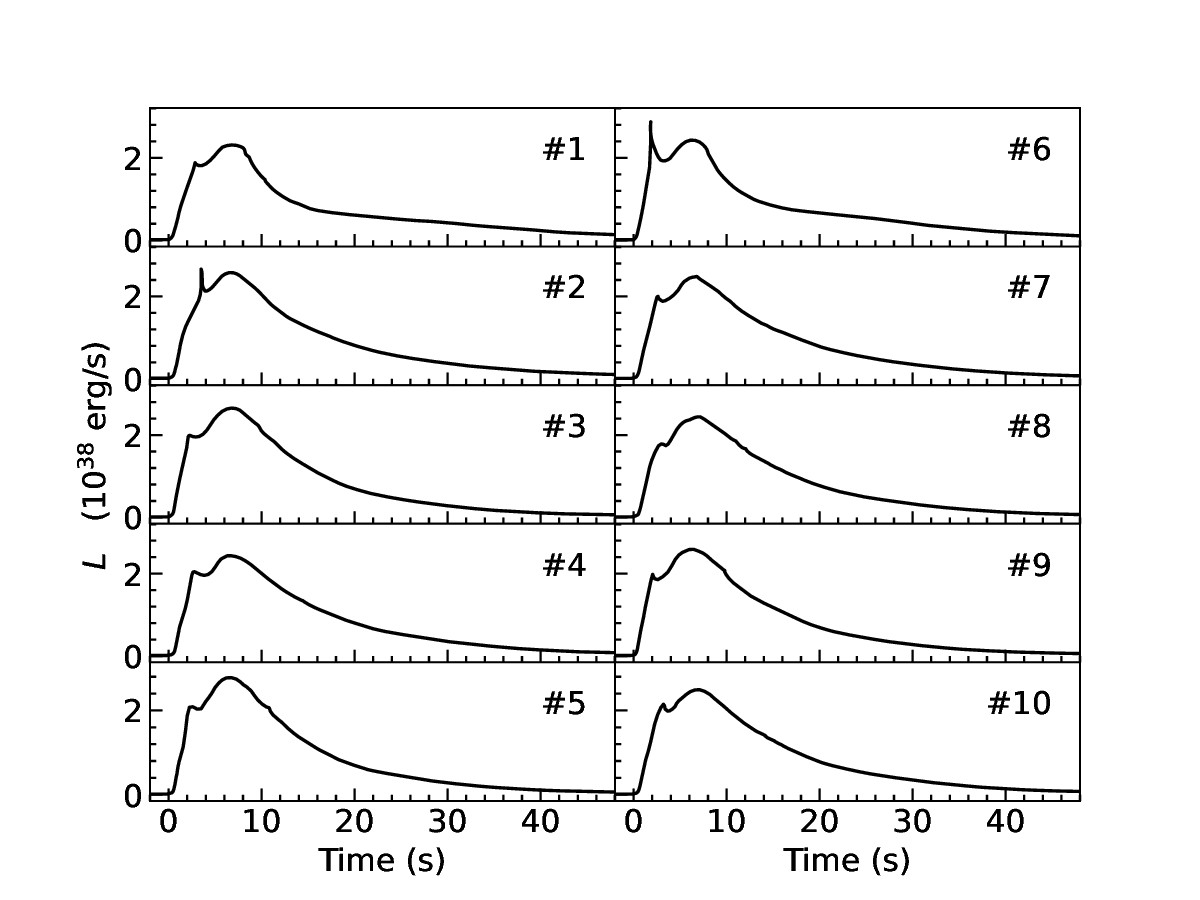}
         \end{minipage} 
         
    \caption{
    Same as Fig.~\ref{fig:ash1} but for standard model 2 with $X=0.6225$, $Y=0.3275$, $Z=0.05$, $\dot{M}=5\times10^{-9}\,M_{\odot}/\rm yr$, $Q_{\rm b}=0.1\,\rm MeV/u$.}
    
    \label{fig:ash2}
\end{figure}
\begin{figure}
    \centering
    \begin{minipage}{1.0\linewidth}
    \centering
         \includegraphics[width=\linewidth]{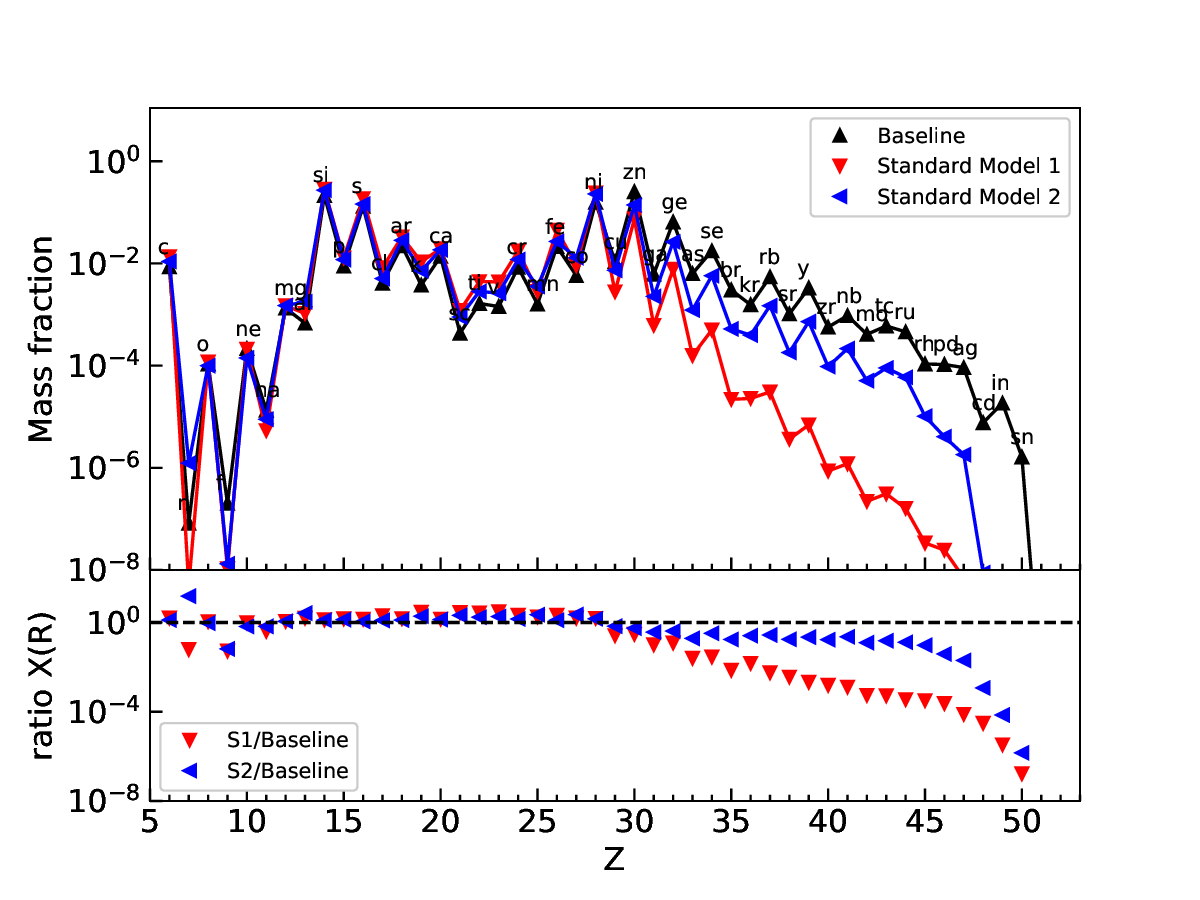}
         \end{minipage} 
         
    \caption{
    The averaged final products for each charge number at the burst tail end with different X-ray burst models. Red: double-peaked bursts from our standard model 1 (S1) with $M=1.4M_{\odot}$, $R=11.2\,\rm km$, $X=0.7325$, $Y=0.2575$, $Z=0.01$, $Q_{\rm b}=0.1\,\rm{MeV/u}$, $\dot{M}=5.0\times 10^{-10}\,M_{\odot}/{\rm yr}$. Blue: double-peaked bursts from our standard model 2 (S2) with $M=1.4M_{\odot}$, $R=11.2\,\rm km$, $X=0.6225$, $Y=0.3275$, $Z=0.05$, $Q_{\rm b}=0.1\,\rm{MeV/u}$, $\dot{M}=5.0\times 10^{-9}\,M_{\odot}/{\rm yr}$. Black: Single-peaked bursts with $M=1.4M_{\odot}$, $R=11.2\,\rm km$, $X=0.70$, $Y=0.28$, $Z=0.02$, $Q_{\rm b}=0.1\,\rm{MeV/u}$, $\dot{M}=1.9\times 10^{-9}\,M_{\odot}/{\rm yr}$, the model is named as baseline model. The lower panel displays the mass faction ratio of the double-peaked bursts to single-peaked bursts.
    }
    \label{fig:ash}
\end{figure}

We discuss the final products during the double-peaked type I X-ray burst nucleosynthesis. The compositional interia due to the leftover of H, He, and CNO nuclei of the previous bursts can have important implications for the properties of subsequent burst~\citep{1980ApJ...241..358T,2004ApJS..151...75W}. In Figs.~\ref{fig:ash1} and~\ref{fig:ash2}, we show a succession of double-peaked bursts for standard models 1 and 2, we can see that the ashes of the previous burst affect the properties of the subsequent burst, which leads to the shape of the double-peaked bursts and final products are different from each burst in one model. Fig.~\ref{fig:ash} shows the averaged final products with standard models 1 and 2. Besides, we calculate the final products during the single-peaked bursts of nucleosynthesis for comparison. As we see, nuclei heavier than $^{56}\rm Ni$ are poorly synthesized for double-peaked bursts, especially for standard model 1 with low mass accretion rate and metallicity. This is because the first peak of a double-peaked burst is caused by a pure helium burst, the amount of pure helium layer of standard model 1 is greater than that in standard model 2, and as a result, the leftover hydrogen of standard model 1 is less than that in standard model 2. Finally, the heavier p-nuclei due to the $rp$ process during the double-peaked bursts nucleosynthesis of standard model 1 is less than that in standard model 2. 

\section{Conclusions}\label{sec:con}
In this work, we investigate the nuclear origins of double-peaked bursts with low mass accretion rate at low metallicity (standard model 1) and one order of magnitude higher mass accretion rate at high metallicity (standard model 2) using MESA. We take into account the variation in mass accretion rate (models 3-10), metallicity (models 11-17), base heating (models 18-23), and nuclear reaction waiting points  (Tables~\ref{tab:nu1}and~\ref{tab:nu2}) to study the properties of the double-peaked bursts. Moreover, we make a comparison between our models and the observations from 4U 1636-53 and 4U 1730-22, the final products during double-peaked bursts nucleosynthesis are also explored. Our conclusions are as follows:

1. As the mass accretion rate increases, the dual-peak luminosity of double-peaked bursts decreases, the first peak time increases, the second peak time remains constant, and the peak separation time decreases. The double-peaked structure disappears at $\dot{M}=9\times10^{-10}\,M_{\odot}/{\rm yr}$ for $Z=0.01$, at $\dot{M}=9\times10^{-9}\,M_{\odot}/{\rm yr}$ for $Z=0.05$, which gives the parameter space for the double-peaked bursts, i.e., $\dot{M}$ should be in the range of $\sim(4-8)\times10^{-10}\,M_{\odot}/{\rm yr}$ for $Z=0.01$, in the range of $\sim(4-8)\times10^{-9}\,M_{\odot}/{\rm yr}$ for $Z=0.05$.

2. As the metallicity increases, the peak luminosity of the first peak decreases, the second peak luminosity changes irregularly, the first peak time decreases, the second peak time remains constant, the peak separation time increases. The double-peaked structure disappears at $Z=0.005$ for $\dot{M}=8\times10^{-10}\,M_{\odot}/{\rm yr}$, at $Z=0.04$ for $\dot{M}=5\times10^{-9}\,M_{\odot}/{\rm yr}$.

3. As the base heating increases, the peak luminosity of the two peaks decreases, both the first and second peak times increase and the peak separation time increases. The double-peaked structure will not disappear with a change of $Q_{\rm b}$ in the range of$\sim0.1-0.4\,\rm MeV/u$. 

4. The uncertainty in the nuclear reaction waiting points: $^{22}\rm{Mg}$, $^{26}\rm{Si}$, $^{30}\rm{S}$, $^{34}\rm{Ar}$, $^{56}{\rm Ni}$, $^{60}\rm Zn$, $^{64}\rm{Ge}$, $^{68}\rm{Se}$, $^{72}\rm{Kr}$ has been explored. We find that $^{22}{\rm Mg}$, $^{56}{\rm Ni}$, $^{64}{\rm Ge}$, $^{68}{\rm Se}$ are possibly the most important nuclear waiting points impedance in the thermonuclear reaction flow to explain the double-peaked bursts.

5. From the model-observation comparison, we find that our nuclear origins of double-peaked bursts can explain the observations such as 4U 1636-53, but it is difficult to explain the observations from 4U 1730-22, which has a large second peak time ($t_{\rm p,2}\approx8.5\,\rm s$)  and peak separation time ($\delta{t}\approx7.5\,\rm s$).

6. The final products during double-peaked bursts of nucleosynthesis are different from that during the single-peaked bursts of nucleosynthesis especially for nuclei heavier than $^{56}\rm Ni$. The heavier p-nuclei due to the $rp$ process during the double-peaked bursts are less than that in single-peaked bursts.

These findings contribute to our comprehension of the diversity observations in double-peaked bursts. However, we can not give a comprehensive parameter space considering more models due to our computational limitation, it is worthwhile for scanning the whole grid of $\dot{M}$ and $z$ parameters to find other pairs of values that would produce double-peaked bursts. As the multipeaked type I X-ray bursts are not observed with regularity, and the distance of the source is uncertain, we do not compare the light curves of the double-peaked bursts but the properties of the double-peaked bursts (e.g., $L_{\rm p,1}$, $L_{\rm p,2}$, $t_{\rm p,1}$, $t_{\rm p,2}$, $\delta t$). Besides, the values of burst strength ($\alpha$), burst energy ($E_{\rm burst}$), recurrence time ($\Delta T$), the first ($L_{\rm p,1}$) and second ($L_{\rm p,2}$) peak luminosity are obtained from our models. However, these values from observations have rarely been studied. The comparisons of these values between the theory and observations are important for our understanding of the nuclear origins of double-peaked bursts, we leave this issue in our future work.

\section*{Acknowledgements}
We appreciate the anonymous referee for helpful comments. We thank M. Hashimoto for his encouragement. We thank W. Wang and A. Dohi for the fruitful discussion. This work received the generous support of the National Natural Science Foundation of China Nos. 12263006, U2031204, 12163005, 12373038, and the Major Science and Technology Program of Xinjiang Uygur Autonomous Region under Grant No. 2022A03013-3.

\section*{Data Availability}
The data underlying this article are available in this article.



\bibliographystyle{mnras}
\bibliography{example} 




\appendix \label{sec:app}

\begin{table*}
    \centering
    \caption{Physical inputs of the double-peaked bursts models and the burst number in each model, where $N_{\rm T}$ represents the total bursts number, $N_{\rm D}$ represents the double-peaked bursts number}.
    \begin{tabular}{ccccccc}
    \toprule
    Standard Model & $\rm X(H)$ & $\rm Y(He) $ & $\rm Z$ & $\rm Q_{b}(MeV/u)$ & $\dot{M}(\rm10^{-9}M_{\odot}/yr)$ & $N_{\rm D}/N_{\rm T}$ \\
    \hline
    1 & 0.7325 & 0.2575 & 0.01 & 0.10 & 0.5 & 24/24\\
    2 & 0.6225 & 0.3275 & 0.05 & 0.10 & 5 & 16/30 \\
    \toprule
    Model Number & $\rm X(H)$ & $\rm Y(He) $ & $\rm Z$ & $\rm Q_{b}(MeV/u)$ & $\dot{M}(\rm10^{-9}M_{\odot}/yr)$  & $N_{\rm D}/N_{\rm T}$ \\
    \hline
    3 & 0.7325 & 0.2575 & 0.01 & 0.1 & 0.4 & 17/17  \\
    4 & 0.7325 & 0.2575 & 0.01 & 0.1 & 0.7 & 11/22 \\ 
    5 & 0.7325 & 0.2575 & 0.01 & 0.1 & 0.8 & 6/21 \\ 
    6 & 0.7325 & 0.2575 & 0.01 & 0.1 & 0.9 & 0/12 \\
    7 & 0.6225 & 0.3275 & 0.05 & 0.1 & 4 & 13/33 \\ 
    8 & 0.6225 & 0.3275 & 0.05 & 0.1 & 7 & 7/30 \\ 
    9 & 0.6225 & 0.3275 & 0.05 & 0.1 & 8 & 6/29 \\
    10 & 0.6225 & 0.3275 & 0.05 & 0.1 & 9 & 0/29 \\
    11 & 0.74625 & 0.24875 & 0.005 & 0.1 & 0.8 & 0/14 \\
    12 & 0.71875 & 0.26625 & 0.015 & 0.1 & 0.8 & 21/24 \\
    13 & 0.705 & 0.275 & 0.02 & 0.1 & 0.8 & 19/25 \\
    14 & 0.65 & 0.31 & 0.04 & 0.1 & 5 & 0/17 \\
    15 & 0.5675 & 0.3625 & 0.07 & 0.1 & 5 & 29/31 \\ 
    16 & 0.54 & 0.38 & 0.08 & 0.1 & 5 & 17/18 \\ 
    17 & 0.485 & 0.415 & 0.10 & 0.1 & 5 & 29/30\\ 
    18 & 0.7325 & 0.2575 & 0.01 & 0.2 & 0.5 & 21/23 \\ 
    19 & 0.7325 & 0.2575 & 0.01 & 0.3 & 0.5 & 23/24 \\
    20 & 0.7325 & 0.2575 & 0.01 & 0.4 & 0.5 & 21/22 \\
    21 & 0.6225 & 0.3275 & 0.05 & 0.2 & 5 & 11/39 \\
    22 & 0.6225 & 0.3275 & 0.05 & 0.3 & 5 & 7/40 \\
    23 & 0.6225 & 0.3275 & 0.05 & 0.4 & 5 & 6/40 \\    
    \hline
    \end{tabular}
    \label{tab:input}
\end{table*}

\begin{table*}
    \centering
    \caption{The output quantities of the double-peaked burst models. The uncertainties of the output values indicate the $1\sigma$ standard deviation.}
    \begin{tabular}{ccccccccc}
    \toprule   
    Standard & $\rm \alpha$ & $\rm E_{burst}$ & $\rm L_{p,1}$ & $\rm L_{p,2}$ & $\rm \Delta T$ & $\rm t_{p,1}$ & $\rm t_{p,2}$ & $\rm \delta t$  \\ 
    Model& $ $ & $10^{39}~\rm erg$ & $10^{38}~\rm erg/s$ & $10^{38}~\rm erg/s$ & $\rm h$ & $\rm s$ & $\rm s$ & $\rm s$  \\
    \hline
1 & $74.95\pm1.82$ & $4.36\pm0.56$ & $2.91\pm0.55$ & $3.37\pm0.59$ & $15.46\pm0.37$ & $1.25\pm0.28$ & $6.47\pm0.89$ & $5.22\pm0.90$\\
2 &  $67.32\pm0.77$ & $3.38\pm0.24$ & $2.06\pm0.26$ & $2.49\pm0.14$ & $1.08\pm0.01$ & $2.85\pm0.43$ & $6.50\pm0.33$ & $3.65\pm0.51$ \\
    \toprule
    Model & $\rm \alpha$ & $\rm E_{burst}$ & $\rm L_{p,1}$ & $\rm L_{p,2}$ & $\rm \Delta T$ & $\rm t_{p,1}$ & $\rm t_{p,2}$ & $\rm \delta t$  \\ 
    Number & $ $ & $10^{39}~\rm erg$ & $10^{38}~\rm erg/s$ & $10^{38}~\rm erg/s$ & $\rm h$ & $\rm s$ & $\rm s$ & $\rm s$  \\
    \hline    
3 & 96.46$\pm2.16$ & 3.74$\pm0.95$ & 5.85$\pm3.27$ & 3.52$\pm1.04$ & 18.79$\pm0.42$ & 0.61$\pm0.34$ & 5.96$\pm0.97$ & 5.35$\pm0.99$ \\
4 & 73.01$\pm0.88$ & 4.58$\pm0.28$ & 2.33$\pm0.75$ & 2.80$\pm0.52$ & 11.57$\pm0.20$ & 1.84$\pm0.35$ & 6.04$\pm0.40$ & 4.20$\pm0.43$ \\ 
5 & 72.07$\pm1.41$ & 4.96$\pm0.46$ & 1.81$\pm0.30$ & 2.77$\pm0.58$ & 10.28$\pm0.28$ & 2.11$\pm0.03$ & 6.24$\pm1.22$ & 4.13$\pm1.27$\\ 
6 & 71.85$\pm2.84$ & 4.71$\pm0.36$ & - & 2.38$\pm0.39$ & 8.89$\pm0.36$ & - & 6.99$\pm0.72$ & -\\ 
7 & 68.25$\pm0.69$ & 3.28$\pm0.16$ & 2.14$\pm0.25$ & 2.51$\pm0.19$ & 1.33$\pm0.01$ & 2.23$\pm0.18$ & 6.45$\pm0.38$ & 4.22$\pm0.44$\\ 
8 & 66.68$\pm1.12$ & 3.53$\pm0.11$ & 1.97$\pm0.16$ & 2.45$\pm0.09$ & 0.79$\pm0.01$ & 2.88$\pm0.26$ & 6.48$\pm0.23$ & 3.60$\pm0.29$\\ 
9 & 65.32$\pm0.88$ & 3.56$\pm0.04$ & 1.84$\pm0.15$ & 2.44$\pm0.08$ & 0.68$\pm0.01$ & 2.90$\pm0.50$ & 6.14$\pm0.61$ & 3.29$\pm0.59$ \\ 
10 & 65.10$\pm1.31$ & 3.59$\pm0.20$ & - & 2.40$\pm0.19$ & 0.62$\pm0.02$ & - & 7.00$\pm0.23$ & - \\ 
11 & 70.57$\pm1.55$ & 5.86$\pm0.80$ & - & 2.21$\pm0.16$ & 12.28$\pm0.26$ & - & 6.83$\pm0.41$ & - \\
12 & 76.45$\pm1.11$ & 3.96$\pm0.43$ & 3.64$\pm1.47$ & 2.79$\pm0.30$ & 8.89$\pm0.13$ & 1.57$\pm0.39$ & 6.19$\pm0.54$ & 4.62$\pm0.62$ \\
13 & 81.83$\pm1.88$ & 3.31$\pm0.32$ & 4.45$\pm1.42$ & 2.93$\pm0.59$ & 7.87$\pm0.18$ & 1.18$\pm0.39$ & 6.55$\pm0.92$ & 5.37$\pm0.84$ \\ 
14 & 66.17$\pm1.08$ & 3.79$\pm0.10$ & 1.61$\pm0.24$ & 2.35$\pm0.11$ & 1.17$\pm0.02$ & 3.12$\pm0.55$ & 6.72$\pm0.44$ & 3.60$\pm0.65$\\  
15 & 72.81$\pm1.45$ & 2.81$\pm0.18$ & 2.66$\pm0.35$ & 2.46$\pm0.35$ & 0.99$\pm0.02$ & 1.62$\pm0.25$ & 6.09$\pm0.53$ & 4.47$\pm0.59$ \\ 
16 & 77.96$\pm1.39$ & 2.54$\pm0.21$ & 3.22$\pm0.52$ & 2.26$\pm0.31$ & 0.95$\pm0.02$ & 1.30$\pm0.25$ & 6.10$\pm0.51$ & 4.80$\pm0.50$ \\ 
17 & 98.75$\pm1.33$ & 1.91$\pm0.22$ & 4.83$\pm0.83$ & 2.24$\pm0.30$ & 0.90$\pm0.02$ & 0.96$\pm0.29$ & 6.29$\pm0.41$ & 5.34$\pm0.43$\\    
18 & 74.32$\pm1.35$ & 4.34$\pm0.28$ & 2.93$\pm0.56$ & 3.31$\pm0.60$ & 15.44$\pm0.27$ & 1.28$\pm0.18$ & 6.60$\pm0.67$ & 5.32$\pm0.65$ \\ 
19 & 74.27$\pm0.81$ & 4.26$\pm0.35$ & 2.96$\pm0.69$ & 3.06$\pm0.59$ & 15.06$\pm0.20$ & 1.29$\pm0.28$ & 6.62$\pm0.67$ & 5.33$\pm0.59$ \\
20 & 73.39$\pm0.70$ & 4.04$\pm0.34$ & 3.21$\pm1.46$ & 2.79$\pm0.67$ & 14.82$\pm0.23$ & 1.31$\pm0.32$ & 6.81$\pm0.83$ & 5.50$\pm0.71$ \\
21 & 67.23$\pm1.03$ & 3.30$\pm0.10$ & 1.86$\pm0.16$ & 2.27$\pm0.18$ & 1.03$\pm0.03$ & 2.95$\pm0.49$ & 6.83$\pm0.60$ & 3.89$\pm0.59$ \\
22 & 65.50$\pm0.45$ & 3.28$\pm0.10$ & 1.65$\pm0.13$ & 2.11$\pm0.14$ & 0.98$\pm0.02$ & 2.99$\pm0.27$ & 7.08$\pm0.15$ & 4.09$\pm0.18$ \\
23 & 62.59$\pm1.27$ & 2.96$\pm0.17$ & 1.40$\pm0.11$ & 1.77$\pm0.07$ & 0.92$\pm0.01$ & 3.66$\pm0.66$ & 8.11$\pm0.54$ & 4.46$\pm0.41$ \\ 
    \hline
    \end{tabular}
    \label{tab:output}
\end{table*}


\bsp	
\label{lastpage}
\end{document}